\UseRawInputEncoding
\documentclass[a4paper,10pt,oneside]{article}
\usepackage[utf8]{inputenc}
\usepackage{icad2023,amsmath,epsfig,times,url,hyperref}
\usepackage[T1]{fontenc}

\hypersetup{breaklinks=true,colorlinks=true,linkcolor=blue,citecolor=blue,urlcolor=blue}

\newcommand{\icadnc}{26}
\newcommand{\icadnp}{1291}

\title{ASTRONOMY \& ASTROPHYSICS IN ICAD HISTORY}

\name{Rub\'en Garc\'ia-Benito} 
\address{Instituto de Astrofísica de Andalucía (IAA), CSIC\\ 
PO Box 3004, 18080 Granada, Spain \\
{\tt rgb@iaa.es}}
\begin{document}
\ninept
\maketitle
\begin{sloppy}
\begin{abstract}

The International Conference on Auditory Display (ICAD) is a significant event for researchers and practitioners interested in exploring the use of sound in conveying information and data. Since its inception in 1994, the conference has served as a vital forum for exchanging ideas and presenting research findings in the field of auditory display. While the conference primarily focuses on auditory display and sound design, astronomy has made its presence felt in the proceedings of the conference over the years. 
 However, its not until the current ICAD conference where astronomy features a dedicated session.
This paper aims to provide an statistical overview of the presence of astronomy in the ICAD conference's history from 1994 to 2022, highlighting some of the contributions made by researchers in this area, as well as the topics of interest that have captured the attention of sound artists. 
\end{abstract}

\section{Introduction}

The International Conference on Auditory Display (ICAD) has a storied history of exploring the boundless possibilities of sound as a means of conveying information and enhancing human interaction with technology. Over the years, this conference has welcomed a diverse community of researchers, practitioners, and artists who have all contributed to the advancement of auditory display. Of particular note is the increasing presence of astronomy-related topics and projects within the ICAD community. This year's conference features a special track dedicated to exploring the intersection of sound and astronomy, which serves as a testament to the growing interest and potential for leveraging sonification as a tool for astronomical research and outreach. In this context, this works aims to give a statistical overview of the presence of astronomy in ICAD's history, beyond the 'classical'\textit{Harmony of the Spheres} theme \cite{Grimm2011}.

\section{METHODOLOGY}

The present investigation utilized an extensive and comprehensive collection of data, consisting of information related to the proceedings of past ICAD conferences spanning the period from 1994 to 2022. To be specific, all individual PDF files available at the Georgia Tech Digital Repository \cite{gtech} were downloaded. As part of the data preprocessing step, six works from the 2006 conference that were found to be duplicates were removed, and the repository was promptly notified, leading to an update in their database. Commencing in 2015, a complete compilation in a single PDF of all ICAD contributions became available, from which individual installations, concerts, and workshops were extracted, since these works are not included in the Georgia Tech Digital Repository. The dataset comprises contributions from 1291 individual works and spans 26 conferences, thereby constituting a robust and representative sample of ICAD conferences.

Although most of the PDF files were text-readable, those from the 1994 conference were scanned copies. To facilitate the analysis of the data, Optical Character Recognition (OCR) software was employed to convert these scanned copies to searchable PDFs. Subsequently, a customized Python program was designed to extract text from each PDF file, match a series of pre-selected keywords, and generate a final table containing individual word counts for each work. By analyzing the full texts of the works, it is possible to identify those that may not explicitly use certain terminology in their titles or abstracts, thereby preventing their exclusion from the analysis.

The present study utilized a pre-selected set of the most popular astronomical terms as keywords to search within the PDF files. These keywords were carefully chosen, and in some cases grouped together, such as 'solar' which includes 'sun', 'stars' which also encompasses 'stellar', and 'galaxies' which includes terms like 'galaxy' and 'galactic'.

Following the selection of the keywords, the resulting word list underwent a manual inspection process to remove words that, while frequent, were not necessarily directly related to astronomy in specific works. Such words included metaphors, loan words, brands, and computer technology terms like 'Sun Microsystems', 'Samsung Galaxy', 'Star Trek', 'Star Wars', 'Supernova digital mixer', among others. There are instances where certain terms share a linguistic root, yet their intended meaning and use may differ. For instance, in \cite{LoPresti1996}, the term 'galaxy' was used in a different context as a borrowed term to describe a large database of documents. Additionally, works that focused on planetary seismology of the Earth or on Earth data were not included in the analysis.

\begin{figure*}[t]
\centering \textbf{Astronomy \& Astrophysics ICAD Word History}
\begin{minipage}[h]{1.0\textwidth}
\centering
\includegraphics[width=1.0\textwidth, trim={0 0 0 1.5cm}, clip]{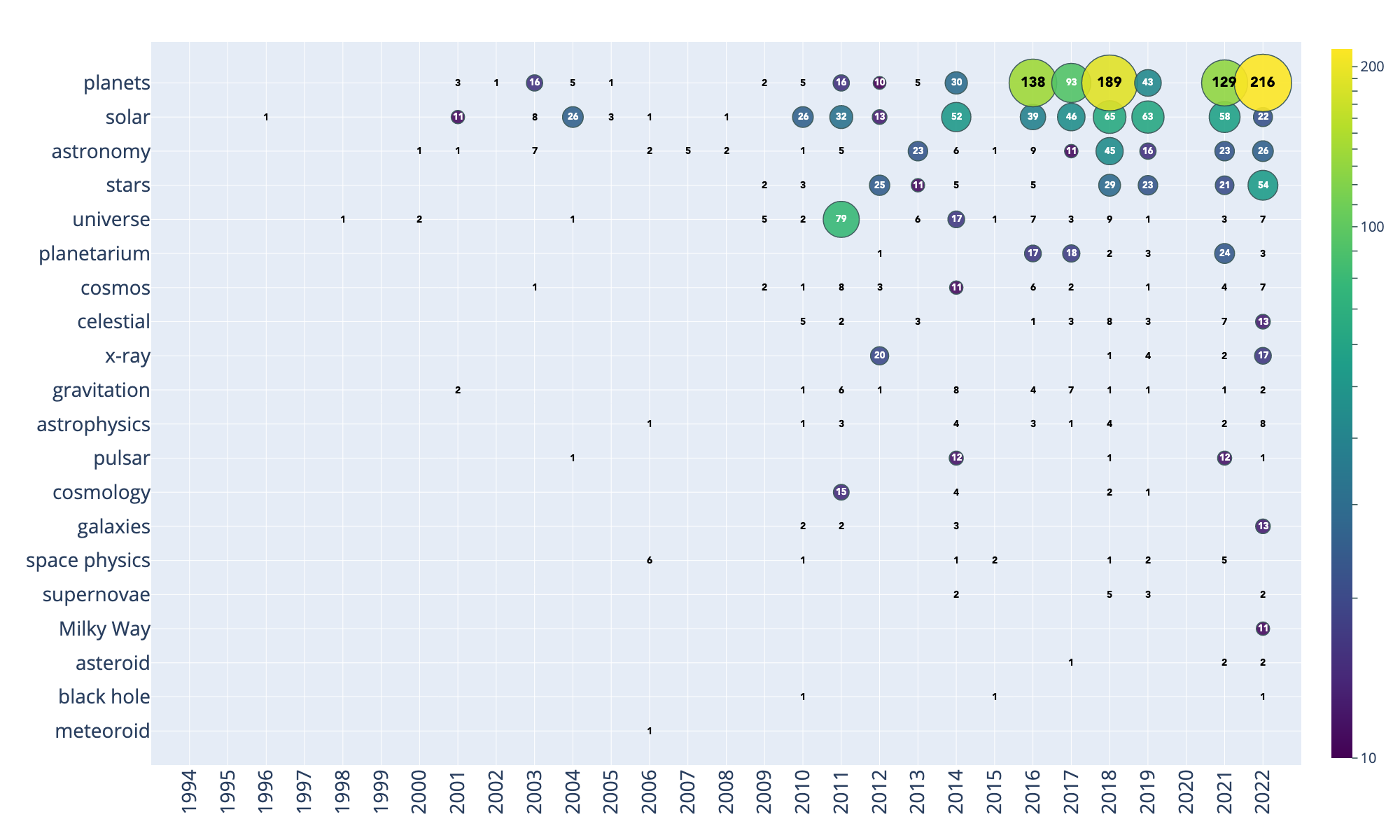}
\end{minipage}

\caption{Bubble plot depicting the frequency of a selection of words related to astronomy in all ICAD proceedings between 1994 and 2022. The selected words are displayed on the vertical axis, while the years are shown on the horizontal axis (data from years 1995, 1999, and 2020 are missing). Each bubble's size and color represent the count number, which is also displayed in the center. For words with less than 10 counts in a given year, only the count number is indicated. The dataset includes contributions from \icadnp\ and spans \icadnc\ conferences. The words on the vertical axis are sorted in descending order based on their overall count throughout the period.}
\label{fig:icad_kws}
\end{figure*}

\section{ASTRONOMY \& ICAD}

Figure \ref{fig:icad_kws} visualizes the frequency of astronomy-related words in ICAD proceedings between 1994 and 2022 using a bubble plot. The selected words are represented on the vertical axis, while the years are displayed on the horizontal axis, with missing data from 1995, 1999, and 2020. Each bubble's size and color represent the count number, which is also displayed in the center. The dataset includes contributions from \icadnp\ and covers \icadnc\ conferences. The words on the vertical axis are sorted in ascending order based on their total count over the entire period.
The outcomes showcased in Figure \ref{fig:icad_kws} don't aim to deliver an exhaustive investigation of the ICAD ecosystem. Rather, the results offer an initial glimpse into the collective sonification mindset and topics of interest, by examining a  sample selected from the most popular astronomical terms. Interestingly, there appears to be a trend towards using the term 'astronomy' more frequently than 'astrophysics' or 'space science'.

Figure \ref{fig:icad_kws} illustrates how astronomy has sparked the imagination of the sound community since the inception of ICAD. While it's true that its utilization has been relatively subdued in the past, the last decade has witnessed an impressive resurgence, particularly from the year 2010 onwards. The upward trajectory of interest is particularly noticeable from 2016, and it has culminated in an exceptional surge in planetary related projects, making it one of the most prominent topic of discussion. The trend has remained relatively stable since then, with the 2022 conference boasting the highest use of 'planets' to date.

There are instances where a particular topic exhibits outlier behavior within a given trend, such as the case of 'universe' in 2011. Most of the counts can be attributed to a single article dedicated to the sonification of the Cosmic Microwave Background \cite{McGee2011}, a collaboration between the departments of physics and music arts of University of California, Santa Barbara.

While sonification is widely transversal and can have broad applications across various fields, there are specific disciplines, such as astronomy, that have invested resources to creating customized and inclusive sonification tools tailored to their data. It is worth noticing the pioneering work by \cite{Candey2006} that aimed to make space physics data accessible to visually-impaired students and researchers by developing a sonification data analysis tool. 

Over time, there have been discernible shifts in the prevailing topic interests. For example, during its early years, the term 'solar' was primarily associated with endeavors that focused on solar radiation or solar wind \cite{Quinn2001,Paine2004,Alexander2011}. 
Around the middle of the 2010s, there was a noticeable increase in the number of sonification projects related to the 'solar system' \cite{Ballora2014}, indicating a growing trend in this area, with some specifically aimed at creating outreach programs for planetarium shows \cite{Quinton2016}, including user evaluation \cite{Elmquist2021}. 

Since 2016, there has been a noticeable increase in interest in planets, as evidenced by various works \cite{Snook2018,Riber2018,Russo2022}, which use astronomy as a source of inspiration for their artistic settings or with an outreach objective. Additionally, some works such as \cite{GarciaRiber2022} have a more analytical user-driven background, and incorporate also user validation as \cite{Quinton2021}.

The focus on the Sun and planets in ICAD history has been significant, with other astronomical topics being less explored. This could be partially attributed to their familiarity with the non-specialist. However, through collaboration with astrophysicists, such as those in \cite{McGee2011}, which can provide access to advanced data in various formats, it is likely for topics in stellar evolution, interstellar medium, extragalactic astronomy or cosmology to emerge and be explored in greater depth in the future.

\section{CONCLUSIONS}

The ICAD community has experienced rapid growth over the course of its nearly three decades in existence, and sonification has garnered interest from a diverse array of fields for its application. A prime illustration of this phenomenon is the current ICAD meeting, which is hosting a special session dedicated to astronomy. While astronomy has long served as a source of inspiration for sound artists and has prompted a multitude of works from both a tool and analysis standpoint, it has only been in the past ten years that its presence has become increasingly prominent, ultimately leading to the establishment of the aforementioned special track. Notably, the principal areas of focus have been the sun, the solar system, and planets. Moving forward, it is reasonable to anticipate that an even broader range of astronomy-related projects will emerge, stemming from various subdisciplines.

To deepen our understanding of the development of sonification in the field of astronomy, the next phase of this research could expand the analysis to include the evolution of topics, trends, and progress in sonification algorithms and methodologies over time. To ensure a more comprehensive overview, it is recommended to move beyond simple word frequency analysis, which may over-represent frequently used words in a single paper. The aim is to create a systematic overview of the field within ICAD history that can serve as a valuable resource for researchers and practitioners alike. By examining the evolution of sonification in astronomy, identifying emerging trends and challenges, and developing new approaches and techniques, this overview could facilitate a deeper understanding of the field and its potential for future advancement.

\section{ACKNOWLEDGMENT}

R.G.B. acknowledges financial support from the grants CEX2021-001131-S funded by MCIN/AEI/10.13039/501100011033,  PID2019-109067-GB100, and to CSIC ``Ayudas de Incorporaci\'on'' grant 202250I003.

\bibliographystyle{IEEEtran}

\bibliography{refs_rgb_ha_icad2023}
%
%
%
%

\end{sloppy}
\end{document}